\documentclass [aps,pra,preprint,showpacs,showkeys,superscriptaddress,]{revtex4}
\usepackage {amssymb}
\usepackage {graphicx}
\usepackage{subfigure}
\usepackage{dcolumn}

\begin{document}

\title{Fusion of necklace-ring patterns into vortex and fundamental solitons in dissipative media}

\author{Y. J. He}
\affiliation{State Key Laboratory of Optoelectronic Materials and
Technologies, Zhongshan (Sun Yat-Sen) University, Guangzhou 510275,
China} \affiliation{School of Electronic and Information
Engineering, Guangdong Polytechnic Normal University, Guangzhou
510665, China}

\author{Boris A. Malomed}
\affiliation{Department of Interdisciplinary Studies, School of
Electrical Engineering,Faculty of Engineering, Tel Aviv University,
Tel Aviv 69978, Israel}

\author{H. Z. Wang}
\thanks{Corresponding author, E-mail: stswhz@mail.sysu.edu.cn}
\affiliation{State Key Laboratory of
Optoelectronic Materials and Technologies, Zhongshan (Sun Yat-Sen)
University, Guangzhou 510275, China}

\begin{abstract}
We demonstrate that necklace-shaped arrays of localized spatial
beams can merge into stable fundamental or vortex solitons in a
generic model of laser cavities, based on the two-dimensional
complex Ginzburg-Landau equation with the cubic-quintic
nonlinearity. The outcome of the fusion is controlled by the number
of "beads" in the initial necklace, $2N$, and its topological
charge, $M$. We predict and confirm by systematic simulations that
the vorticity of the emerging soliton is $|N-M|$. Threshold
characteristics of the fusion are found and explained too. If the
initial radius of the array ($R_0$) is too large, it simply keeps
the necklace shape (if $R_0$ is somewhat smaller, the necklace
features a partial fusion), while, if $R_0$ is too small, the array
disappears.
\end{abstract}



\maketitle

\section{INTRODUCTION}
The complex Ginzburg-Landau (CGL) equation is a universal model
which plays an important role in many areas such as in
superconductivity and superfluidity, fluid dynamics,
reaction-diffusion phenomena, nonlinear optics, Bose-Einstein
condensation, and quantum field theories \cite{Aranson02, Rosanov02,
Malomed05}. The CGL models produce dynamical behaviors, e.g.,
spatiotemporal chaos and formation of periodic patterns to
dissipative solitons and their bound states \cite{Aranson02,
Rosanov02, Malomed05, Akhmediev05, Malomed91}.

Many recent works focused on localized complex patterns in
conservative models of optical media, different from the simplest
nodeless ground-state modes, such as vortex solitons \cite{ZChen05},
soliton clusters \cite{Desyatnikov02,Kartashov02}, dipole-mode
structures and their multipole counterparts
\cite{Garcia-Ripoll00,Neshev01,Carmon01,Kartashov05,YJHe06}, and
necklace-ring solitons (NRSs)
\cite{Solja¡¦cic¡ä98,Mihalache03,Yang05}. Similar patterns may be
stable in dissipative models based on the CGL equation with the
cubic-quintic (CQ) nonlinearity. These include stable localized
vortices in two- and three-dimensional (2D and 3D) CGL equation
\cite{Crasovan01,Mihalache06,Mihalache07} and necklace-shaped
soliton clusters \cite{Skryabin02}. Recently, stable spatiotemporal
NRSs were also reported in the 3D CQ-CGL equation \cite{He06}.

Previous studies of complex dissipative-solitons patterns were
focused on their stability. A more general issue is a possibility of
dynamical transformation (fusion) of an initial soliton cluster into
complex structures of a different type, or into a simple fundamental
soliton. In this work, we demonstrate that necklace patterns can
fuse into stable fundamental and vortex dissipative solitons in the
2D CQ complex GL equation, which is controlled by topological
numbers of the initial pattern. Such an outcome is only possible in
dissipative models, where the energy and angular momentum are not
dynamical invariants. We predict the vorticity of the eventual
state, and corroborate it by numerical results. If the initial
radius of the necklace array is too large, it features a partial
fusion, or simply keeps the initial structure; on the other hand, if
the initial radius is too small, the pattern decays to zero.

\section{THE MODEL}
Following the notation adopted in Ref. \cite{Crasovan01}, we
consider the CGL equation with the CQ nonlinearity in the
(2+1)-dimensional setting, with propagation distance $Z$ and
transverse coordinates, $X$ and $Y$:

\begin{eqnarray}
 iu_Z+i{\alpha}u+(1/2-i\beta)(u_{XX}+u_{YY})+(1-i\varepsilon)|u|^2u-(\nu-i\mu)|u|^4u=0, \label{Eq:1}
\end{eqnarray}

where $\alpha > 0$ and $\mu> 0$ are parameters of the linear and
quintic loss (the latter one accounts for the gain saturation),
$\varepsilon > 0$ is the cubic gain, $\nu$ (that may be both
positive and negative) accounts for the quintic
self-defocusing/focusing, the diffraction and cubic self-focusing
coefficients are  normalized to be 1, and $\beta > 0$ is the
effective diffusion coefficient. Except for the latter one, all
other coefficients are standard ingredients of optical models based
on the CGL equations; as for $\beta$, it appears in models of laser
cavities, as \cite{Lega94}
$\beta=-\tau_{p}\tau_{c}\Delta/(\tau_{p}+\tau_{c})^{2}$ (in
normalized units), where $\tau_{p}$ and $\tau_{c}$ are the
polarization-dephasing and cavity-decay times, and $\Delta$ detuning
between the cavity's and atomic frequencies. Therefore, the relevant
case of $\beta > 0$ corresponds to the negative detuning.

We stress that the CQ nonlinearity in Eq. (1) is not a truncated
expansion of a saturable nonlinearity, but represents a fundamental
response of the nonlinear medium, due to some intrinsic resonance(s)
in it. This response was directly observed in chalcogenide glasses
\cite{Smektala00,Boudebs03} and in some organic optical materials
\cite{Zhan02}, although saturable nonlinearity is more generic in
optics \cite{Rosanov02}.

The model based on the CGL equation (1) does not take into regard
the finite relaxation time of the gain and loss in optical media,
which may be added to the model through the respective evolution
equation \cite{Rosanov02,Rosanov06,Fedorov07}. However, it seems
quite plausible that the results reported below will not be strongly
affected by this modification of the model.

It has been demonstrated that vortex solitons and soliton clusters
can self-trap in the framework of the 2D CQ-CGL model (and its 3D
extension), provided that coefficient $\beta$ is presented
\cite{Crasovan01,Mihalache06,Mihalache07,Skryabin02,He06}. The
stability of the localized patterns is supported by the simultaneous
balance between the transverse diffraction and self-focusing, and
between the gain and linear and quintic losses (including the
effective diffusive loss, accounted for by $\beta > 0$).

Following Refs. \cite{Solja¡¦cic¡ä98,Mihalache03,Yang05}, the
initial necklace-shaped pattern, with amplitude $A$, mean radius
$R_0$, and width $w$, can be taken (in polar coordinates $r$ and
$\theta$) as

\begin{eqnarray}
\label{Eq:2}
  u(Z=0,r,\theta)=Asech[(r-R_0)/w]\cos(N\theta)exp(iM\theta),
  \end{eqnarray}
Here, integer $N$ determines the number of elements ("beads") in the
ring (necklace) structure, which is $2N$, while another integer,
$M$, is the topological charge of the complex pattern.

\section{Numerical results}

Analysis of numerical results demonstrates that generic outcomes of
the evolution of the necklace array can be displayed, e.g., for
$\alpha = 0.5$, $\beta = 0.5$, $\varepsilon= 2.5$, $\nu= 0.01$, and
$\mu= 1$, which corresponds to a physically realistic situation and,
simultaneously, makes the evolution relatively fast, thus helping to
elucidate its salient features
\cite{Mihalache03,Mihalache06,Mihalache07}. In this case, the
amplitude and width of the individual 2D stable fundamental soliton,
as found from Eq. (1), are $A = 1.6$ and $w = 1$. The robustness of
the emerging patterns was tested in direct simulations of Eq. (1)
with the initial condition taken as expression (2) additionally
multiplied by $[1+\rho(x)]$ where $\rho(x)$ is a Gaussian random
function with zero average, the mean size of the perturbation
amounting to 10 percent of the soliton's amplitude.

The array with the topological charge equal to half the number of
"beads" in the necklace, i.e., $M=N$, merges into a fundamental
soliton, under the condition that initial radius $R_0$ of the
necklace array is smaller than a certain critical value,
$R_{max}^{F}$, see Fig. 1(a). As follows from Eq. (2), the mean
phase shift between adjacent beads in the initial array is
$\Delta{\varphi} = \Delta{\varphi}_0 + 2{\pi}M/(2N) = 0$, where
$\Delta{\varphi_{0}=-\pi}$ corresponds to the opposite sign of
adjacent "beads" in the necklace with $M = 0$. Therefore, individual
elements in the array, being in-phase, attract each other, which
leads to their fusion into a stable fundamental soliton, as seen in
Figs. 1(b, c).

Note that the CGL equation, being a dissipative one, does not have
any dynamical invariant, hence the total angular momentum is not
conserved either. In fact, the initial configuration corresponding
to Eq. (2) with $M = N$ may be realized as a mixture of two states,
with values of the vorticity (total topological charge)
$S=M-N\equiv{0}$ and $S=M+N$. Obviously, the effective diffusion
term in underlying Eq. (1), which is proportional to $\beta$,
produces a much stronger dissipation effect on the component of the
wave field with the nonzero vorticity, hence only the zero-vorticity
one survives in the limit of $Z\rightarrow\infty$, as observed in
Fig. 1.

It is also easy to understand the increase of $R_{max}^{F}$ with
$N$, see Fig. 1(a). Indeed, the attraction between adjacent "beads",
necessary for their fusion into the fundamental soliton, is not too
weak if the separation between them does not exceed a certain
(maximum) value \cite{Malomed91}. The radius of the necklace,
corresponding to a given separation between the "beads", grows
linearly with $N$, which explains the roughly linear form of
dependence $R_{max}^{F}(N)$in Fig. 1(a).

 When radius $R_0$ of the initial necklace pattern is larger than a certain minimum value,
$R_{min}^{V}$, but smaller than $¡Ö1.8N$, i.e.,
\begin{eqnarray}
\label{Eq:3}
 R_{min}^{V}\leq{R_0}\leq{1.8N},
  \end{eqnarray}
the array with topological charge $M$ evolves into a stable vortex
soliton, provided that $M$ falls into either of two mutually
symmetric intervals, $M_{min}^{V}\leq{M}<N$ or
$N<M<M_{max}^{V}\leq{2N-M_{min}^{V}}$ see Fig. 2. As said above,
initial configuration (2) may be realized as a mixture of states
with vorticities whose absolute values are $S=|N\pm{M}|$, hence the
vorticity component which may survive in the course of the evolution
(one which is least affected by the diffusion term) is
$S_{fin}=|N-M|$. Indeed, the simulations confirm that the emerging
vortex soliton features precisely this value of the vorticity.
Because the asymptotic form of the vortex at $r\rightarrow 0$ is
const$\cdot{r^{S}}$, smaller values of $S\equiv{|N-M|}$ correspond
to a smaller radius of the inner hole in the vortex soliton, as
observed in Figs. 2(c-e). Dependence $M_{min}^{V}(R_0)$, which is
displayed in Fig. 2(b) for fixed values of $N$, can be explained
too. Indeed, well-pronounced minima in the dependence at
intermediate values of $R_0$ corresponds to the fact that the fusion
of the necklace array into vortex soliton is most feasible when the
initial radius of the pattern is close to the radius of the expected
vortex ring, which determines the location of the minima in Fig.
2(b).

On the other hand, when the initial radius of the necklace is
smaller than the minimum value, $R_{min}^{V}$ [see Eq. (3)], the
pattern rapidly disappears (decays to zero) upon the propagation,
irrespective of the value of $M$ , as shown in Fig. 2(f). This
effect is explained by the strong dissipation generated by the
effective diffusion term in Eq. (1) (the one proportional to
$\beta$) in the necklace array of a small radius. Naturally, the
diffusive dissipation is stronger for a larger azimuthal gradient.
The latter grows, on the average, linearly with $ N$ , that is why
$R_{min}^{V}$ also increases with $ N$ , as seen in Fig. 2(a).

If the initial radius of the necklace exceeds the above-mentioned
maximum values, i.e., for the case of $M = N$, and $\approx{1.8N}$
for $M\neq{N}$, as per Eq. (3) (recall these two cases correspond to
the fusion of the necklace into the fundamental and vortex soliton,
respectively), Fig. 3 shows that the necklace arrays with M = 0 do
not undergo the fusion, but feature slow expansion, while their
counterparts with $M\neq{0}$ make the number of the "beads" smaller
via the fusion involving some of them, after passing several hundred
diffraction lengths. The number of the lost "beads" is larger for
smaller values of $|M - N|$, and smaller initial radius of the
necklace, $R_0$.

Finally, if the initial radius exceeds a still larger threshold
value, $R_{min}^{N}$, the interaction between the "beads" in the
necklace array becomes negligible, irrespective of its topological
charge M (which implies $R_{min}^{N}$ is defined with some
uncertainty, as it may slightly vary with the change of the total
propagation distance). As a result, the pattern keeps its
necklace-like structure and the initial radius, as shown in Fig. 4.
For the same reasons as mentioned above [cf. Fig. 1(a)], the
dependence of the respective minimum radius $R_{min}^{N}$ on
modulation number $N$ of initial pattern (2) is approximately
linear, see Fig. 4(a). Note that each individual element in the
necklace patterns observed in Figs. 4 features an isotropic
(circular) shape, unlike the "beads" in the initial pattern. This is
explained by the fact that each element evolves into a fundamental
soliton, whose deformation induced by the interaction with its
neighbors is negligible, since the distance to them is too large. In
fact, the "freezing" of soliton necklaces of a large radius due to
the exponential decay of the interaction forces was observed in many
cases before \cite{Solja¡¦cic¡ä98 ,Mihalache03,Yang05}.

\section{CONCLUSION}
We have demonstrated that, unlike conservative systems described by
multidimensional nonlinear Schr{\"{o}}dinger equations, models of
dissipative optical media (in particular, laser cavities
\cite{Lega94}), based on the 2D CGL equation with the cubic-quintic
nonlinearity, admit fusion of necklace-ring patterns into stable
fundamental and vortical solitons, provided that the initial radius
of the ring, $R_0$, is not too large (if $R_0$ is somewhat larger,
the necklace may undergo a partial fusion). The outcome of the
evolution is controlled by values of modulation number N and
topological charge $M$ of initial necklace (2). A simple analysis
has predicted the vorticity of the fused soliton to be $|N - M|$,
which is fully confirmed by direct simulations. Threshold
characteristics of the fusion process, such as those displayed in
Figs. 1(a) and 2(a, b), were also explained in a qualitative form.
The possibility to control the outcome of the fusion by means of the
initial topological numbers, $N$ and $M$, suggests a principal
possibility to use the process in all-optical switching schemes. On
the other hand, if the initial radius of the necklace array is too
large, the interaction between individual elements is negligible,
allowing the pattern to keep its initial radius and necklace
structure, while each element assumes the isotropic shape,
corresponding to fundamental dissipative solitons in 2D.

In addition to the straightforward realizations in terms of
nonlinear optics, the model may also find application to
Bose-Einstein condensates, where the dissipation and gain occur in
models of matter-wave lasers, as demonstrated experimentally and
theoretically in various settings
\cite{Malomed98,Kneer98,Miesner98,Hagley99,Drummond99,Schneble03,Carr04,Rodas-Verde05,Chen05,Carpentier06}.
Thus far, such models were analyzed only in the effectively 1D
geometry, while the present results suggest to extend them into two
dimensions.

\section{acknowledgements}
This work was supported by the National Natural Science Foundation
of China (10674183) and National 973 Project of China
(2004CB719804), and Ph. D. Degrees Foundation of Ministry of
Education of China (20060558068).

\newpage
\begin{figure}
\vspace{3cm}
  \centering
   \includegraphics[totalheight=6.5cm,width=8.5cm]{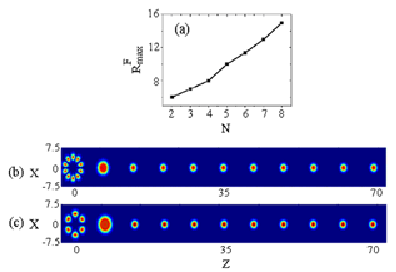}
   \vspace{1.5cm}
  \caption{\label{fig:1}(Color online) The fusion into a fundamental soliton of the
necklace array whose initial radius is not too large,
$R_{0}\leq{R_{amx}^{F}}$ , and the topological charge is equal to
half the number of "beads" in the array, $M = N$. (a): The largest
radius, admitting the fusion, versus N. (b, c): Examples of the
fusion for $R_{0} = 4$ and $M = N = 5$ (b) or $M = N = 3$ (c) [the
examples are shown by means of contour plots of the local power,
$|u(x,y)|^{2}$ ].}
\end{figure}

\newpage
\begin{figure}
\vspace{3cm}
  \centering
   \includegraphics[totalheight=13cm,width=8.6cm]{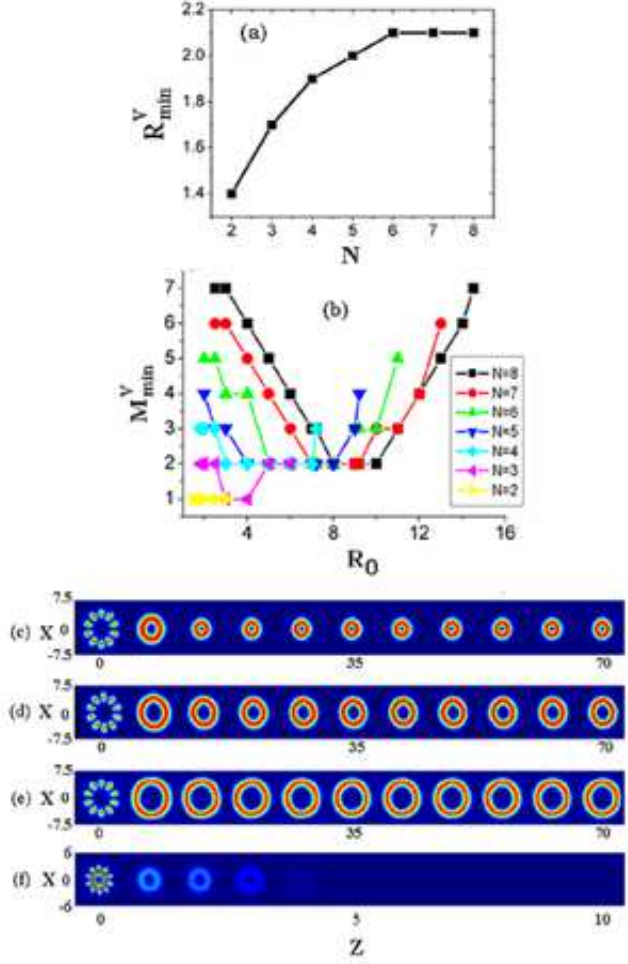}
   \vspace{1.5cm}
  \caption{\label{fig:2}(Color online) The fusion into a stable vortex soliton of
the necklace whose initial radius and topological charge satisfy
constraints  $R_{min}^{V}\leq{R_0}\leq{1.8N}$, and either
$M_{min}^{V}\leq{M}<N$ or ${N<M<M_{max}^{V}}\leq{2N-M_{min}^{V}}$.
(a): The minimum initial radius admitting the fusion versus $N$;
(b): the minimum topological charge versus the initial radius for
different fixed values of modulation number $N$. (c-e): Examples of
the formation of the vortex soliton with $N = 5$ and  $R_{0} = 5$
for $M = 4$ (c), $M = 3$ (d), and $M = 2$ (e). Additionally, panel
(f) displays an example of the decay of the necklace cluster with $N
= 5$ and $M = 4$, in the case of $R_{0}<R_{min}^{V}$ (here,
$R_{0}=1.9$ and $R_{min}^{V}=2$).}
\end{figure}

\newpage
\begin{figure}
\vspace{3cm}
 \centering
  \includegraphics[totalheight=3.3cm,width=8.6cm]{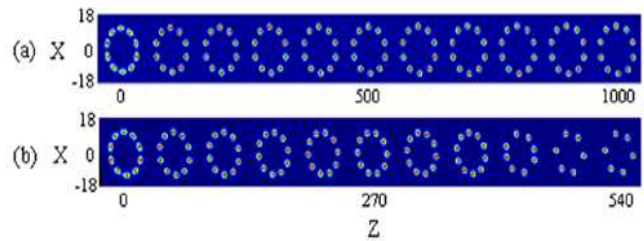}
    \vspace{1.5cm}
 \caption{\label{fig:3} (Color online) (a): Slow expansion of the necklace array with
$N = 5$, $M = 0$ and $R_{0} = 11$, in the case when its initial
radius slightly exceeds the maximum value $1.8N$, see Eq. (3). (b):
In the same case, but with $M = 3$, ten initial "beads" fuse into
eight and, eventually, into six individual elements.}
\end{figure}

\newpage
\begin{figure}
\vspace{3cm}
 \centering
  \includegraphics[totalheight=8.2cm,width=8.4cm]{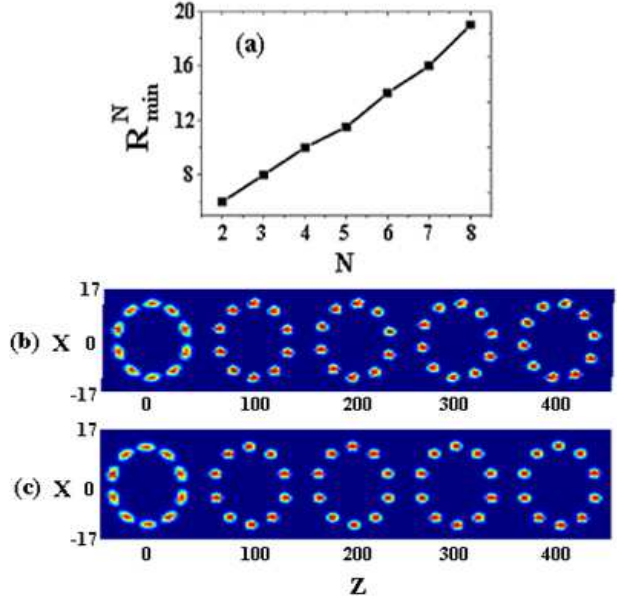} \hspace{0.00cm}
    \vspace{1.5cm}
    \caption{\label{fig:4} (Color online) Formation of "frozen" patterns which look
like stable soliton necklaces, with the initial necklace radius
exceeding $R_{min}^{N}$ and an arbitrary value of the topological
charge (M). (a): $R_{min}^{N}$ as a function of modulation number
$N$ (b, c): Examples of the formation of stable necklace rings with
$N = 5$ and $R_{0} = 12$ for $M = 2$ (b) and $M = 0$ (c).}
\end{figure}

\end{document}